\documentclass[pdflatex,sn-mathphys-num,iicol]{sn-jnl}

\usepackage{graphicx}%
\usepackage{multirow}%
\usepackage{amsmath,amssymb,amsfonts}%
\usepackage{amsthm}%
\usepackage{mathrsfs}%
\usepackage[title]{appendix}%
\usepackage{xcolor}%
\usepackage{textcomp}%
\usepackage{manyfoot}%
\usepackage{booktabs}%
\usepackage{algorithm}%
\usepackage{algorithmicx}%
\usepackage{algpseudocode}%
\usepackage{listings}%

\theoremstyle{thmstyleone}%
\theoremstyle{thmstyletwo}%

\theoremstyle{thmstylethree}%

\begin{document}

\title{Control and Navigation of a 2-D Electric Rocket}


\author{\fnm{André} \sur{Fonte}}\email{\{andrevfonte, pedrodossantos31, paulo.j.oliveira\}@tecnico.ulisboa.pt}

\author{\fnm{Pedro} \sur{dos Santos}}

\author{\fnm{Paulo} \sur{Oliveira}}

\affil{\orgname{ISR/IST and IDMEC/IST}, \orgaddress{\street{Av. Rovisco Pais 1}, \city{Lisbon}, \postcode{1049-001}, \country{Portugal}}}


\abstract{This work addresses the control and navigation of a simulated two-dimensional electric rocket. The model provides a simplified framework that neglects actuator dynamics and aerodynamic effects while capturing the complexities of underactuation and state coupling. 
Trajectory tracking is achieved through a modularized and layered control architecture, with employement of a Linear Quadratic Regulator (LQR) and Lyapunov theory.
Full-state estimation is achieved through Kalman filtering techniques, part of the navigation module.
The solutions are thoroughly evaluated in a custom-built MATLAB/Simulink testbed, simulating real-world conditions
while maintaining a simplified setup. The results reveal limitations along the lateral axis, whose resolution is suggested for future work.}

\keywords{UAV, Control, Navigation, Kalman Filter, LQR, Lyapunov}



\maketitle

\section{Introduction}\label{sec:intro}

The recent years have seen the ever-growing space industry quickly evolve, which
has driven the development of advanced and efficient control and navigation solutions.
These systems must address complex dynamics of attitude and position while
compensating for external disturbances, mostly stochastic.
Real-world testing of these systems is both challenging and expensive.
As such, extensive simulation testing has become an essential tool for
minimizing costs and ensuring reliable performance.
Studies similar to those of~\cite[Spannagl et al.]{spannagl2021}
and~\cite[Linsen et al.]{linsen2022} highlight the effectiveness of using inexpensive
prototypes as a testbed for testing control and navigation systems.
Rockets are often underactuated, meaning the arity of independent
control inputs is less than the number of degrees of freedom.
Space industries adapt this constrained design in the name of lighter
vehicles and increased efficiency despite the higher complexity of the derivation of control laws~\cite{brocket2017}.
This work focuses on addressing these challenges within the context of a small-scale,
two-dimensional electric rocket (referred to as e-rocket), described by a simplified yet nonlinear model that encapsulates underactuation and state coupling effects.
The proposed approach prioritizes simplicity and clarity, similarly to the first stages of industry or academically-oriented projects.

This study begins with the six-state nonlinear dynamics of the e-rocket, from which nonlinear and linear control solutions are derived.
Linearization procedures are common within the aerospace industry as they enable the
employment of well-known solutions to complex rocket~\cite{santos2025pitch} or other vehicles~\cite{martins:inner-outer-feedback2022}, as well as the
tackling of underactuated systems through full-state
feedback~\cite{madeiras:fullstate2024}.
As a Jacobian-linearized model retains its validity only in the vicinity of the trimming conditions, only the attitude control is designed based on a purely linear model, thus enabling feedback control solutions with employment of Linear Quadratic Regulators (LQR) with integral action.
This inherent limitation to classical linear control justifies the derivation of Lyapunov-based modules por position regulation, which aim to ensure the stability of the e-rocket for any magnitude of initial deviations.
The stochasticity of real-world scenarios is mimicked through artificial injection of disturbances, which are compensated by the navigation modules, based on Kalman filtering principles.

These theoretically-sound systems are developed based on a common underlying targeted trajectory, described by a set of waypoints that the e-rocket must reach.
The proposed trajectory consists of a strictly vertical ascending movement with constant climbing velocity, operated at an arbitrarily-defined horizontal position.
All control modules are validated through simulations in a custom-built MATLAB/Simulink environment, which aims to assess the performance of the approach for the proposed trajectory.

This work is structured as follows: Section~\ref{sec:modeling} addresses the mathematical
model of the e-rocket and its underlying simplifications;
Section~\ref{sec:control} presents the control system design, including the linear and nonlinear control solutions, briefly assessing its performance individually;
Section~\ref{sec:navigation} details the navigation subsystems, targeting the estimation of the corrupted states;
the simulation setup and 2-D tracking results are presented in Section~\ref{sec:simulation};
Finally, some concluding remarks are presented in Section~\ref{sec:conclusions}.

\section{Model Description}\label{sec:modeling}

\begin{figure}
    \centering
    \includegraphics[width=.65\linewidth]{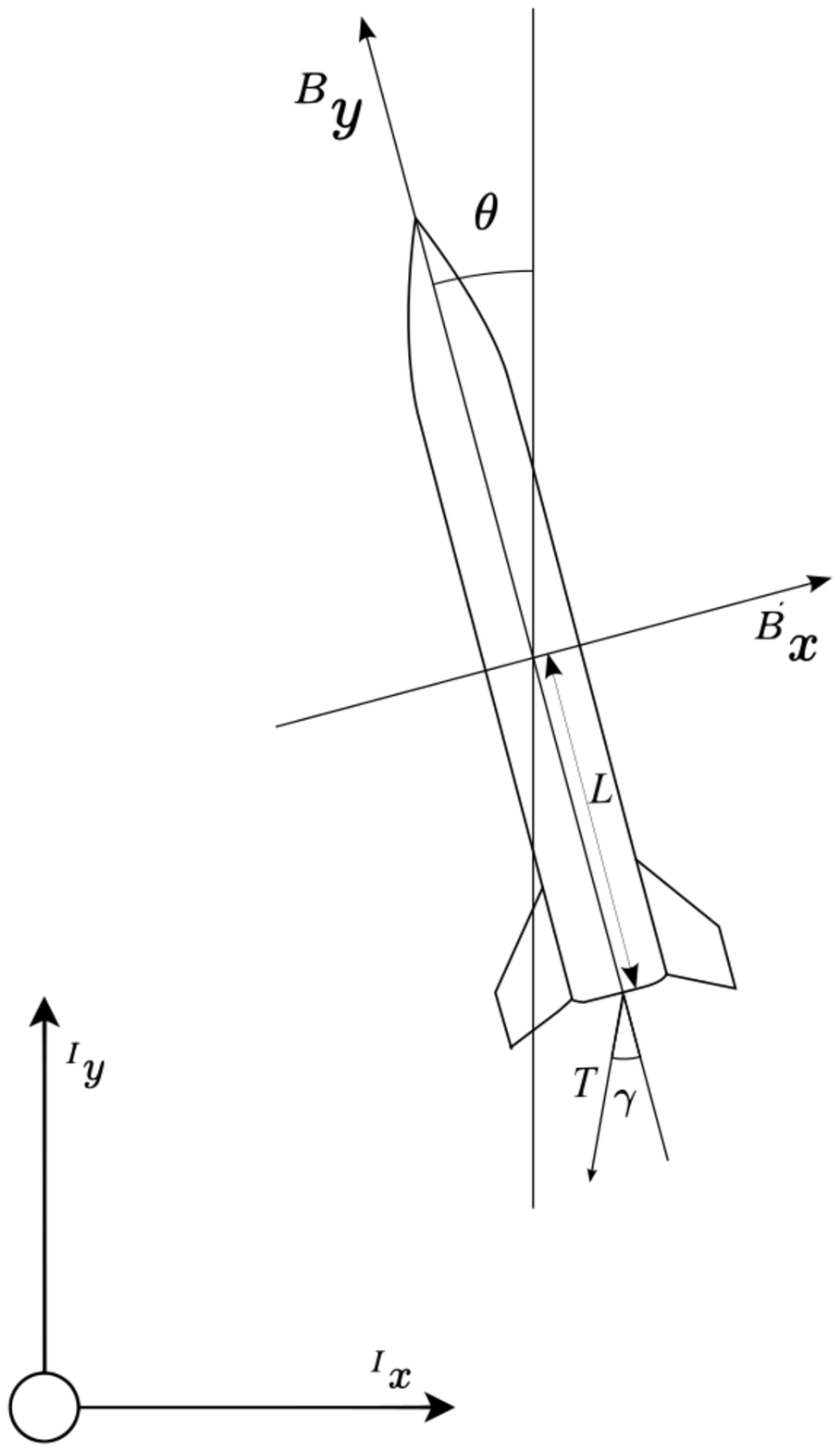}
    \caption{Two-dimensional representation of the e-rocket.}
    \label{fig:rocket_draw}
\end{figure}

The two-dimensional model for the kinematics and dynamics of the e-rocket is defined
with respect to the inertial reference frame $\{I\}$, with origin at the ground, and a body reference frame $\{B\}$ centred at the centre of mass of the prototype.
Using both these references, converted into one another through a set of rotations,
facilitates the perception and description of the rocket's motion.
For simplicity, the equations assume negligible aerodynamic and actuator dynamics.
The two-dimensional model of the e-rocket, as represented in Fig.~\ref{fig:rocket_draw}, is obtained using Newton-Euler's equations~\cite{santos2023}, yielding
\begin{align}
    &\mathbf{\dot{p}}=\mathbf{Rv} \label{eq:1} \\
    &\mathbf{\dot{R}}=\mathbf{RS}(\omega) \label{eq:2} \\
    &m\mathbf{\dot{v}}=-m\mathbf{S}(\omega)\mathbf{v}-mg\mathbf{R^T}\mathbf{e_y}
    +\mathbf{f_T} \label{eq:3} \\
    &J\dot{\omega}=\tau \label{eq:4}
\end{align}
where
$\mathbf{{p}} = \begin{bmatrix}
    x&y
\end{bmatrix}^T$
is the rocket's position in $\{I\}$,
$\mathbf{{v}} = \begin{bmatrix}
    u&v
\end{bmatrix}^T$
is its velocity in $\{B\}$, $\omega$ is its angular velocity with respect to
the normal of the plane of motion,
$\mathbf{{R}} = \begin{bmatrix}
    \cos{\theta}&-\sin{\theta}\\
    \sin{\theta}&\cos{\theta}
\end{bmatrix}$
is the 2-D rotation matrix from $\{B\}$ to $\{I\}$---defined as
function of the Euler pitch angle $\theta$---and
$\mathbf{{f}_T} = \begin{bmatrix}
    T\sin{\gamma}&T\cos{\gamma}
\end{bmatrix}^T$
is the 2-D thrust vector, dependent on the deflection angle $\gamma$ of the actuation and on its magnitude $T$.
As such, the model considers thrust vector actuation through angular deviations on the thrusting force, applied at $L$ meters below the centre of mass of the body.
Also function of the thrust and tilt angle is the torque, given by $\tau=LT\sin\gamma$.
The skew-symmetric matrix of the angular speed, $\omega$, is denoted by $\mathbf{S}
(\omega) =
\begin{bmatrix}0&-\omega
\\ \omega&0
\end{bmatrix}$.
Finally, $m$ and $J$ represent the mass and moment of inertia of the rocket.
For the present work, a small-scale rocket is considered with the form of a solid
cylinder.
With a mass of $m=2$ kg, an arm of $L=0.5$m, a total length of $L_b=1.5$m, and a radius $r\ll L_b$, its moment
of inertia along the considered rotation axis is approximated to
$J=mL_b^2/12=0.3750 \text{ kg}\text{m}^{-2}$.
Gravitational acceleration is considered constant and equal to $g=9.81$ m/s$^2$.

The set of equations from~\eqref{eq:1} to~\eqref{eq:4} comprehensively describe how the rocket's position, velocity and orientation evolve over time under external forces, design parameters and control inputs.
Equation~\eqref{eq:1} tracks the rocket's translational kinematics in space as its
body-frame velocity changes.
Similarly, equation~\eqref{eq:2} outlines rotational kinematics, translating the
angular rate into changes in orientation.
The translational dynamics of the e-rocket are described by equation~\eqref{eq:3};
velocity is considered to be under the influence of---from left to right---inertia,
gravity and the thrust vector.
Equation~\eqref{eq:4} captures the rotational dynamics,
detailing the impact of the thrust vector on the angular velocity of
the rocket.
The thrust vectoring system is key
in enabling the control of both translational and rotational motion, and its effects
are idealized to be immediate.
By combining these four equations, we can establish the rationale for mediating the
interplay between the thrust vector actuator (TVA) and the rocket's behaviour.

A six-equation system can be trivially derived from the presented equations, resulting in
\begin{equation}
    \begin{cases}
        \dot{x} = u\cos\theta - v\sin\theta \\
        \dot{u} = v\omega -g\sin\theta + \frac{T}{m}\sin\gamma \\
        \dot{y} = u\sin\theta + v\cos\theta \\
        \dot{v} = -u\omega - g\cos{\theta} + \frac{T}{m}\cos\gamma \\
        \dot{\theta} = \omega\\
        \dot{\omega} = \frac{LT}{J}\sin{\gamma}
    \end{cases}\label{eq:model-6eq-system},
\end{equation}
an explicit yet less compact representation that eases out the understanding of the
coupling between states $(x,u,y,v,\theta,\omega)$ and control inputs $(T,\gamma)$.
This notation highlights the underactuated nature of the system, which adds particular challenges to the motion control in free space.

\section{Control}\label{sec:control}

The following section proposes a modularized architecture combining linear and nonlinear approaches to tackle the states' regulation according to the desired trajectory.

From the six-state model in~\eqref{eq:model-6eq-system}, it's trivial to realize greater influence of each of the TVA parameters on each set of states, given the proposed trajectory, foreshadowing two modes of motion for the e-rocket.
This statement is verifiable through the Jacobian linearization of the full model, considering the trimming conditions for a steady vertical flight:
\begin{equation*}
    \begin{aligned}    
        &\mathbf{x}^* = \begin{bmatrix}
                           x_d & 0 & \dot{y}_d\cdot t & \dot{y}_d & 0 & 0
        \end{bmatrix}^T,\quad
        &\mathbf{u}^* = \begin{bmatrix}
                           mg & 0
        \end{bmatrix}^T\label{eq:op-point-vertical-flight}
    \end{aligned}
\end{equation*}
where $\mathbf{x}^*$ and $\mathbf{u}^*$ are the nominal values for the states and inputs, respectively. 
The targeted horizontal position is $x_d$, the desired velocity is $\dot{y}_d$, and time is denoted by $t$.
Under these conditions, the lateral mode---described by $\delta x$, $\delta u$, $\delta \theta$, and $\delta \omega$---is acted upon by the deflection of the thrust vector, while the longitudinal mode---described by $\delta y$ and $\delta v$---is influenced by the thrust magnitude.
The prefix $\delta$ denotes the deviation from the nominal values.

Independent modules are developed to each set of states, simplifying the design and stability analysis of the control system.
This is a common practice in the industry, as modularity provides isolation of effects, prevents propagation of errors, and eases out their debugging.
Specifically, the altitude controller ensures soft convergence to the targeted altitude, relying on thrust magnitude adjustments.
Simultaneously, an attitude controller adjusts the rocket's intended orientation---determined by an outer position tracker---through the deflection of the thrust vector, ensuring the e-rocket maintains a vertical trajectory at the desired horizontal position. 

\begin{figure}[ht]
    \centering
    \includegraphics[width=.9\linewidth]{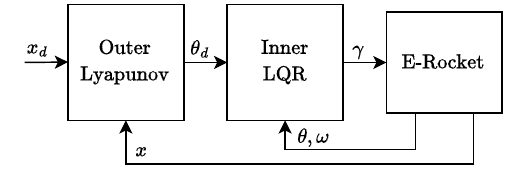}
    \caption{Inner-outer loop layout for horizontal convergence.}
    \label{fig:inner-outer.pdf}
\end{figure}

The setup is architectured such that global convergence is guaranteed position-wise, i.e., the e-rocket is able to reach any set of input waypoints, regardless of the magnitude of the initial deviations.
The second method of Lyapunov provides this guarantee.
The close coupling between rotational and transverse dynamics is addressed by an inner-outer loop structure, where the inner loop rapidly tracks the pitch angle references determined by the outer law though linear techniques, as represented in Fig.~\ref{fig:inner-outer.pdf}.
On the other hand, altitude tracking is achieved through a backstepping-based solution that ensures the system reaches the desired altitude at the correct time, as dictated by the reference trajectory.

\subsection{Attitude Regulation}\label{subsec:attitude-tracking}

The deductions within the following subsections (\ref{subsec:horizontal-position-tracking} and~\ref{subsec:vertical-position-tracking}) are based on the assumption of a fully operational attitude controller that continuously tracks and adjusts the e-rocket's rotation through the TVA deflection. The design of this controller is presented in the current section.
The nonlinear rotational dynamics presented in~\eqref{eq:model-6eq-system} are trivially linearized around the trimming conditions for a steady vertical flight, i.e., $\theta^* = \omega^* = \gamma^* = 0$ and $T^* = mg$.
The linear equivalent of the rotational dynamics is then described by a standard double integrator.

Integral action introduces memory to the closed-loop system, enforcing it to correct for steady-state errors.
Its incorporation is operated through the addition of a new state to the two-state rotation model, $\zeta_\theta$, quantifying the integrated pitch tracking error, i.e.,
\begin{equation*}
    \dot{\zeta}_\theta = \delta\mathbf{\theta}_d - \delta\theta
\end{equation*}
thus extending the open-loop model to the triple integrator below, represented in state-space form:
\begin{equation}
    \begin{aligned}
        &\begin{bmatrix}
            \delta\dot{\theta} \\ \delta \dot{\omega} \\ \dot{\zeta}_\theta
        \end{bmatrix}
        = \begin{bmatrix}
            0&1&0\\0&0&0\\-1&0&0
        \end{bmatrix}
         \begin{bmatrix}
            \delta{\theta} \\ \delta {\omega} \\ {\zeta}_\theta
        \end{bmatrix}
        + \begin{bmatrix}
            0\\Lmg/J\\0
        \end{bmatrix}\delta \gamma\\
        &\delta \theta = \begin{bmatrix}
                1 & 0 & 0
        \end{bmatrix}\mathbf{x}\label{eq:3state_atti}
    \end{aligned}
\end{equation}
where the extended states vector is $\mathbf{{x}}=\begin{bmatrix}
    \delta \theta & \delta \omega & \zeta_{\theta}
\end{bmatrix}^T$, the input variable is $u=\delta \gamma$ and the pitch assumed available.
The classical controllability matrix, defined for linear time-invariant systems has rank 3, determining that the simplified system described in~\eqref{eq:3state_atti} is fully controllable.
Fully controllable linear systems inherently satisfy the Brockett's condition~\cite{brocket2017}, ensuring the existence of a smooth state feedback law that stabilizes the linearized plant.
The closed-loop dynamics are then governable by a feedback gain matrix $\mathbf{K} = \begin{bmatrix}
    k_p & k_d & -k_i
\end{bmatrix}$,  where $k_p, k_d, k_i > 0$, employed as
\begin{equation*}
    \delta\gamma = -\begin{bmatrix}
    k_p & k_d
\end{bmatrix}\begin{bmatrix}
        \delta{\theta} \\ \delta {\omega}
    \end{bmatrix} + {k}_i\zeta_\theta
\end{equation*}
which balances control stiffness with the magnitude of the deviations in a biased manner.
Cost-minimization techniques such as the Linear Quadratic Regulator (LQR) can be employed to determine the optimal values for the feedback gains.
The LQR method is based on the minimization of a quadratic cost function, defined as
\begin{equation*}
    J = \int_{0}^{+\infty} \left( \mathbf{x}^T\mathbf{Q}\mathbf{x} +
    \mathbf{u}^T\mathbf{R}\mathbf{u} \right) dt
\end{equation*}
where $\mathbf{Q}$ and $\mathbf{R}$ are the state cost matrix and input cost matrix,
respectively.
These matrices quantify the relative penalties for state deviations and control efforts,
defining the system's notion of optimality.

The correspondent steady-state gain matrix is given by
$\mathbf{K}=\mathbf{R}^{-1}\mathbf{B}^T\mathbf{P}$,
where $\mathbf{P}$ is the positive-definite matrix that
satisfies the continuous-time Riccati equation:
\begin{equation*}
    \mathbf{A}^{T}\mathbf{P} + \mathbf{P}\mathbf{A} -
    \mathbf{P}\mathbf{B}\mathbf{R}^{-1}\mathbf{B}^T\mathbf{P} + \mathbf{Q} =
    \mathbf{0}
\end{equation*}
which depends on the system and on the cost matrices.
The selection of $\mathbf{Q}$ and $\mathbf{R}$ is critical and often challenging in
real-world scenarios, as it depends heavily on the system and environmental dynamics and
constraints.
Fine-tuning these matrices requires a deep understanding of the mission objectives,
system behaviour, and operational environment.

Considering the scope of this work, the matrices were adjusted based on the controller's performance when tracking a step reference.
In the end, the covariance matrices were deemed to be $\mathbf{Q} = \texttt{diag}(100, 5, 1000)$ and $R = 100$.
From these values results the following control law:
\begin{equation}
    \delta \gamma = - \begin{bmatrix}
        1.9558 & 0.4467 & -3.1623
    \end{bmatrix}\begin{bmatrix}
        \delta \theta \\ \delta \omega \\ \zeta_\theta
    \end{bmatrix}.
    \label{eq:control_law_gamma}
\end{equation}

\begin{figure}[ht]
    \centering
    \includegraphics{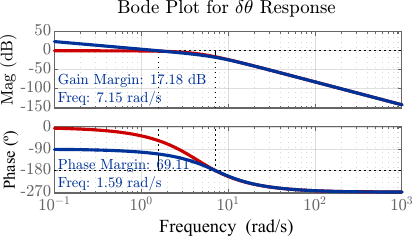}
    \caption{Bode plot of (in red) closed- and (in blue) open-loop pitch dynamics with controlling action.}
    \label{fig:bode_theta.pdf}
\end{figure}

Implementing the regulator just designed into the simplified plant model closes the loop and yields new dynamics, which can be analysed in both the time and frequency domains.
The Bode plot of the closed-loop pitch dynamics is presented in Fig.~\ref{fig:bode_theta.pdf}, outlining the low-frequency pass-band and high-frequency attenuation.
The gain margin of $17.18$ dB for the open-loop transfer function verifies the closed-loop stability.

\begin{figure}[ht]
    \centering
    \includegraphics{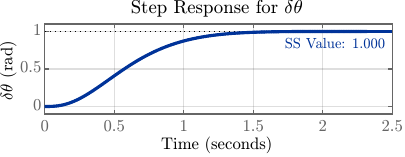}
    \caption{Step response of closed-loop pitch dynamics.}
    \label{fig:step_theta.pdf}
\end{figure}

The flat unitary gain for low frequencies is a direct consequence of the integral action, which ensures that the steady-state error is zero, even when faced with disturbances with non-zero mean.
The step response of the closed-loop pitch dynamics outlines the accurate tracking of the pitch angle given a constant reference, being displayed in Fig.~\ref{fig:step_theta.pdf}.
The system rapidly converges to the desired pitch angle with no overshoot, ticking all the boxes for a well-designed controller.

\subsection{Horizontal Position Tracking}\label{subsec:horizontal-position-tracking}

To design the horizontal position regulator, we assume a fully operational thrust controller that maintains the rocket at a constant body velocity equal to the trajectory reference $\dot{y}_d$;
this subsystem is designed in Section~\ref{subsec:vertical-position-tracking}.
The objective is to continuously adjust a pitch reference such that the vertical motion is operated at an arbitrarily-defined horizontal position, $x_d$, deemed constant within the scope of this work.

The horizontal kinematics of the e-rocket can be simplified by assuming driftless motion, i.e., $u\approx 0$.
This is a drastic yet common approximation, often used in early-stage aerospace projects.
It results in
\begin{equation}
    \dot{x} = -\dot{y}_d\sin\theta
    \label{eq:simplified_horizontal_kinematics}
\end{equation} 
when considering the body velocity to be $v=\dot{y}_d$, which translates direct control over the horizontal position to the pitch angle.
Through its manipulation, the tracking error in $x(t)$, described as $\tilde{x}(t)=x(t)-x_d$, can be driven to zero.
The closed-loop dynamics of the inner loop are fast enough such that the outer component sees the pitch angle adjustments as nearly immediate, i.e., $\theta \approx \theta_d$. 

Let us consider the candidate Lyapunov function (CLF) 
\begin{equation*}
    \mathcal{V}(\tilde{x}) = \frac{1}{2}\tilde{x}^2
\end{equation*}
which is continuous, positive definite, and has continuous partial derivatives.
Its time derivative is
\begin{equation*}
    \dot{\mathcal{V}} = \tilde{x}\dot{\tilde{x}} = \tilde{x}\dot{x} = -\tilde{x}\cdot \dot{y}_d\sin\theta = -k_x \tilde{x}^2,\quad k_x>0,
\end{equation*}
when considering $
    k_x\tilde{x}=\dot{y}_d\sin\theta,$ where $k_x$ is a strictly positive tuning gain, constrained to the interval $[0,\dot{y}_d/\tilde{x}]$.
The negative definiteness of the time derivative ensures the asymptotical stability of the system.
The Global Invariant Set Theorem extends the stability to a global scale, as the largest set of points where $\dot{\mathcal{V}} = 0$ is the origin, i.e., $\tilde{x}=0$.
The pitch reference is then determined by the outer loop as
\begin{equation}
    \theta_d=\arcsin{\left(\frac{k_x}{\dot{y}_d}\tilde{x}\right)}
    \label{eq:guidance_theta}
\end{equation}
allowing for adjustments in the pitch angle based on the error in the horizontal position.
The e-rocket should be able to maintain a strictly vertical trajectory after converging to the desired horizontal position.

The performance of the presented inner-outer loop structure is validated through custom-built simulations, where the plant module is described by the considered kinematics in~\eqref{eq:simplified_horizontal_kinematics} and the attitude-related equations in~\eqref{eq:model-6eq-system} (fifth and sixth equations).
These are subject to the control laws of~\eqref{eq:control_law_gamma} and~\eqref{eq:guidance_theta}, connected as described by Fig.~\ref{fig:inner-outer.pdf}.

\begin{figure*}[ht]
    \centering
    \includegraphics{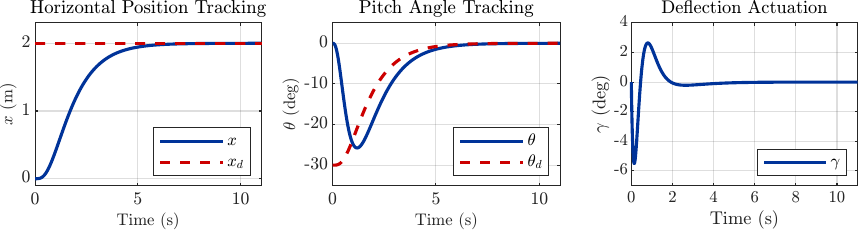}
    \caption{Lateral motion when tracking a 2-meter reference for the horizontal position with simplified kinematics.}
    \label{fig:horizontal_response.pdf}
\end{figure*}

The tracking results for a 2-meter step reference for the horizontal position are presented in Fig.~\ref{fig:horizontal_response.pdf}, where the horizontal position is shown to converge to the desired value by actuating over the pitch angle.
Both modules of the horizontal control operate harmoniously, effectively aligning the rocket's body velocity with the vertical axis.
The presented results were obtained for a tuning gain of $k_x=0.5$ and a body velocity of $\dot{y}_d = 2$ m/s.

\subsection{Altitude Tracking}\label{subsec:vertical-position-tracking}

The vertical position regulator is designed to maintain the rocket such that the rocket converges to a desired inertial altitude, $y_d(t)$, expected to vary with time.
This time-varying scenario justifies the employment of a more robust control solution, such as the Lyapunov-based backstepping method, which ensures the convergence of both the tracking error and its derivative to the origin.
Note that if the ascending trajectory were enforced solely through position convergence, the control solution would be simpler but prone to an uncorrectable delay, causing the rocket to climb at the correct velocity but with a slight lag.
This method presents recursive Lyapunov functions, hence the simultaneous stabilization of the tracking error and its derivative.

Let us define an auxiliary states vector $\mathbf{x}_\text{in}$ as
\begin{equation*}
    \mathbf{x}_\text{in} = \begin{bmatrix}
        \tilde{y} \\ \dot{\tilde{y}}\end{bmatrix}
= \begin{bmatrix}
    y-y_d \\ \dot{y}-\dot{y}_d \end{bmatrix}
= \begin{bmatrix}
    \alpha_1 \\ \alpha_2
\end{bmatrix}
\end{equation*}
where $\tilde{y}$ quantifies the deviation of the state from the desired altitude $y_d$.
The time derivative of $\mathbf{x}_\text{in}$ is fully defined once the second derivative of the tracking error is determined.
The result can be verified to be 
\begin{equation*}
    \dot{\mathbf{x}}_\text{in} = \begin{bmatrix}
        \dot{\tilde{y}} \\ \ddot{\tilde{y}}
    \end{bmatrix} =
    \begin{bmatrix}
        u\sin\theta + v\cos\theta -\dot{y}_d \\ \frac{T}{m} \cos{(\gamma-\theta) - g}
    \end{bmatrix}
    = \begin{bmatrix}
        \dot{\alpha}_1 \\ \dot{\alpha}_2
    \end{bmatrix}
\end{equation*}
for $\ddot{y}_d$ equal to zero.

To first ensure the stabilization of the lower-order state, an auxiliary CLF is defined as
\begin{equation*}
    \mathcal{V}_1(\alpha_1) = \frac{1}{2}\alpha_1^2,
\end{equation*}
which is provable to be continuous, positive definite, and with continuous partial derivatives.
The time derivative of the CLF is then
\begin{equation*}
    \dot{\mathcal{V}}_1 = \alpha_1\dot{\alpha}_1 = \alpha_1\alpha_2 = -k_1\alpha_1^2,\quad k_1>0,
\end{equation*}
which, when considering a positive tuning gain $k_1$, is negative definite, ensuring asymptotical stability. 
The Global Invariant Set Theorem ensures that said stability is global since that largest set of points where $\dot{\mathcal{V}}_1 = 0$ is the origin, i.e., $\alpha_1=0$.
The time-derivative state $\alpha_2$ acts as the stabilizing virtual input, regulated as $(\alpha_2)_d = -k_1\alpha_1$, thus implying $(\dot{\alpha}_2)_d = -k_1\alpha_2$.

The second-order state is then stabilized by a second CLF definition, written as
\begin{equation*}
    \mathcal{V}_2(\alpha_1,\tilde{\alpha}_2) = \mathcal{V}_1(\alpha_1)+\frac{1}{2}\tilde{\alpha}_2^2,
\end{equation*}
where $\tilde{\alpha}_2 = \alpha_2 - (\alpha_2)_d$, which retains the same properties as the previous CLF.
The time derivative of the second CLF is
\begin{equation*}
    \begin{aligned}
    \dot{\mathcal{V}}_2 &= -k_1 \alpha_1^2 +  \tilde{\alpha}_2 \dot{\tilde{\alpha}}_2 \\
    &=  -k_1 \alpha_1^2 + \tilde{\alpha}_2 \left(  \frac{T}{m} \cos{(\gamma-\theta)} - g +k_1\alpha_2 \right) \\
    &= -k_1 \alpha_1^2 - k_2 \tilde{\alpha}_2^2,\quad k_2 > 0,
    \end{aligned}
\end{equation*}
whose negative definiteness ensures the asymptotical stability of the second-order state, extended to global stability by the Global Invariant Set Theorem.
Under these conditions, the thrust magnitude is fully determined by
\begin{equation*}
    T = m\frac{g-k_1 \alpha_2 - k_2 \tilde{\alpha}_2}{\cos{(\gamma - \theta)}}
\end{equation*}
where we remember that $\alpha_2 = \dot{\tilde{y}}$ and $\tilde{\alpha}_2 = \dot{\tilde{y}} + k_1\tilde{y}$.

\begin{figure*}[ht]
    \centering
    \includegraphics{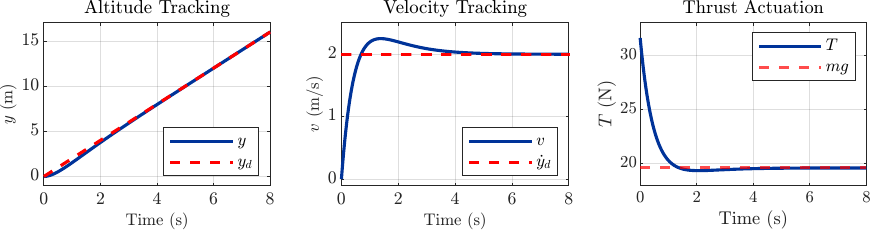}
    \caption{Vertical motion while tracking a time-varying altitude reference increasing at 2 m/s with no lateral motion.}
    \label{fig:vertical_response.pdf}
\end{figure*}

The performance of the proposed solution is evaluated through simulations, where the plant is described by the nonlinear altitude kinematics and dynamics, as considered for the just described control law.
Once again, the gain tuning process resulted in the values $k_1=2$ and $k_2=1$.
The tracking results for a 2-meter per second climb are presented in Fig.~\ref{fig:vertical_response.pdf}, where the altitude is shown to converge to the desired values through the depicted actuation over the thrust magnitude.
A slight overshoot is observed in the evolution of the body velocity as an effort of the controller to ``catch up'' with the desired trajectory in a timely manner after an initial delay.

\section{Navigation}\label{sec:navigation}

In previous sections, the control solutions were obtained assuming full
knowledge of all relevant states.
However, real-world scenarios are inherently affected by noise and uncertainty in the state
measurements.
The foreseen navigation system aims to minimize the propagation of these
inaccuracies to the control components, optimizing the state estimation through sensor fusion techniques.

A set of motion sensors to be installed on-board such as GPS, accelerometers, rate gyros, and inclinometers are simulated to provide noisy measurements.
Kalman filtering is employed for state estimation, which benefits
from its stochastic formulation,
minimizing the mean squared error of estimating quantities corrupted by zero-mean
Gaussian noise.
Combining complementary sensor outputs enhances robustness of the state estimation,
as it combines the strengths of each sensor.

Two independent navigation systems are developed for estimating the altitude and attitude states.
In order not to fall into redundant deductions, the horizontal position and lateral velocity are assumed to be perfectly known, meaning they are not subject to navigation considerations.

\subsection{Attitude Filter}\label{subsec:attitude-filter}

The rotation kinematics of the e-rocket are given by
\begin{equation*}
    \begin{aligned}
    &\dot{\theta}(t) = \omega_m(t) - \eta(t) \\
    &\theta_m(t) = \theta(t) + \xi(t)
    \end{aligned}
\end{equation*}
where the rate gyros provides measurements for the angular velocity
of the rocket w.r.t.\ the inertial frame, $\omega_m(t)$, and the inclinometer provides
pitch measurements, $\theta_m(t)$.
Both measurements are affected by zero-mean white noise, where  $\eta(t)\sim N(0,q)$
represents the process noise and $\xi(t)\sim N(0,r)$ the measurement noise.
The process noise $\eta(t)$ impacts the angular rate,
being integrated over time and thereby introducing a drift to the pitch estimate, while
the inclinometer's noise $\xi(t)$ induces high-frequency fluctuations.

Translating the system into state-space notation allows for a systematic application
of Kalman theory.
The filter is designed based on the plant model and on the noise covariance matrices,
which determine the Kalman gain vector, $\mathbf{L}$.
Its computation involves solving the Riccati equation, given by
\begin{equation*}
\mathbf{PA}^T+\mathbf{AP}-\mathbf{PC}^T
\mathbf{R}^{-1}\mathbf{CP}+\mathbf{Q}=\mathbf{0}
\end{equation*}
in order to $\mathbf{P}$, the covariance of the estimation error matrix.
Then, the steady-state gains are obtained as
$\mathbf{L}=\mathbf{PC}^T\mathbf{R}^{-1}$.

Applying the described theory to the attitude stochastic system yields the
single-entry matrices
\begin{equation}
    p=\sqrt{qr},\quad l=\sqrt{q/r}\label{eq:covariance_matrices_attitude}
\end{equation}
whose formulation highlights implications on the filter design.
The gain $l$ is adjusted based on the process-to-measurement noise ratio, optimally
estimating the pitch angle by weighting the measurements from the angular rate sensor
and the inclinometer.
A higher $q/r$ ratio indicates greater uncertainty in the state evolution, leading to an
increase of the Kalman gain $l$;
this adjustment causes the filter to rely more on the inclinometer measurements for
pitch estimation.
For a low-quality inclinometer, strongly disturbed by measurement noise, the rate gyros
measurements are given higher trust by reducing the gain.
Implementing the estimator into the rotation kinematics model yields
\begin{equation}
    \dot{\hat{\theta}}=\omega_m+l(\theta_m-\hat{\theta}).
    \label{eq:attitude_estimator}
\end{equation}

The sensor fusion performed by the filter exhibits properties that justify its
classification as a complementary filter.
The rate gyros captures rapid variations to the rocket's pitch but leads to drift
over time due to accumulation of process noise on the integrated angle;
it's the other way around for the inclinometer measurements, which are reliable in the
long-term even if noisier around the expected value.
The combination of both is implemented through complementary filtering, where the rate
gyros
output is processed with a high-pass, for short-term accuracy, and the inclinometer data
filtered with a low-pass filter, for accurate convergence over time.
Considering $F_\theta(s)$ and $F_\omega(s)$ to be the transfer functions from the
measured quantities ($\theta_m$ and $\omega_m$, respectively) to the estimated pitch
$\hat{\theta}$ written in the $s$-domain, the plant of~\eqref{eq:attitude_estimator}
determines that
\begin{equation*}
    F_{\theta}(s) = \frac{l}{s+l},\quad
    F_{\omega}(s) = \frac{1}{s+l}
\end{equation*}
which satisfies the relation $F_{\theta}(s) + sF_{\omega}(s) = 1$.
Taking $s \rightarrow 0$ or $s\rightarrow \infty$ demonstrates the low or high-pass
characteristics of the addition terms.

Referring back the Kalman gain computation
of~\eqref{eq:covariance_matrices_attitude}, the noise covariance matrices were set to $q =
10^{-6}\,[(\text{rad/s})^2]$ and $r = 10^{-6}\,[\text{rad}^2]$, yielding a Kalman gain
of $l=1$ which balances the weights of both sensors equally.
The suitability of these values depends on the noise sources, which can be defined within
the simulation environment to match these attributes.
The resulting closed-loop system has a single negative real eigenvalue equal to $-l=-1$.

The effectiveness of the estimators can be assessed through injection of noisy
quantities into the filters, as conducted in Section~\ref{sec:simulation}, with inclusion of the controllers previously designed.
As simulations are conduced with a fixed-step solver (the Euler solver in particular), the noise power parameter of the \texttt{Band-Limited White Noise} block is numerically equal to product between the covariances fed into the Kalman filters and the sampling time. This conclusion is directly derived from the considerations of~\cite[Vasconcelos (2013)]{vasconcelos2013}.

\begin{figure}[ht]
    \centering
    \includegraphics{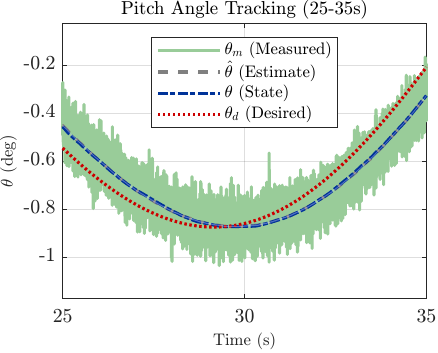}
    \caption{Pitch angle estimation through the Kalman filter.}
    \label{fig:pitch_angle_estimation.pdf}
\end{figure}

A 10-second interval of the pitch angle estimation, is presented here in Fig.~\ref{fig:pitch_angle_estimation.pdf}.
The filter is shown to accurately track the pitch angle in the mist of the noise, as depicted by the blatant overlap between the estimated and true values.
The attenuation of the high-frequency noise is verifiable through the low magnitude of the standard deviation, computed to be $0.0024$ degrees.

\subsection{Altitude Filter}\label{subsec:position-filter}

Regarding the vertical navigation sensors, an accelerometer and a GPS
are considered to be installed on-board, providing measurements for the vertical
acceleration, $a_m$, and position, $y_m$.
These quantities are corrupted by additive white noise, where yet again $\eta(t)\sim N(0,q)$
represents the process noise and $\xi(t)\sim N(0,r)$ stands for the measurement noise.
The described system can be translated as
\begin{equation*}
    \begin{aligned}
    &\dot{y}(t) = v(t) \\
    &\dot{v}(t) = a_m(t) - \eta(t) \\
    &y_m(t) = y(t) + \xi(t)
    \end{aligned}
\end{equation*}
or, in state space notation, as
\begin{equation*}
    \begin{aligned}
    &\begin{bmatrix} \dot{y}(t) \\ \dot{v}(t) \end{bmatrix} = \begin{bmatrix}
        0&1\\0&0
    \end{bmatrix}\begin{bmatrix} y(t)\\v(t) \end{bmatrix} + \begin{bmatrix}
        0\\1
    \end{bmatrix}a_m(t)-\begin{bmatrix}
        0\\1
    \end{bmatrix}\eta(t)\\
    &y_m(t)=\begin{bmatrix}
        1&0
    \end{bmatrix}\begin{bmatrix} y(t)\\v(t) \end{bmatrix}+\xi(t)
    \end{aligned}.
\end{equation*}

The Kalman filter finds the optimal estimates based on the
transmitted energy from the noise sources.
After solving the Riccati equation in order to $\mathbf{P}$, the Kalman gain matrix is
determined as function of the noise covariance matrices, simplifying to
\begin{equation*}
    \mathbf{L} = \begin{bmatrix}
                     \sqrt {2} \sqrt[4]{q/r} \\ \sqrt{q/r}
    \end{bmatrix}\label{eq:kalman-gain-position-estimator}
\end{equation*}
for the stated system.
The $q/r$ ratio directly influences the balance between the two sensor measurements.
When the accelerometer readings are highly disturbed, a larger process noise $q$ causes
the Kalman gains to increase, thus shifting greater trust to the GPS position
measurements.
Conversely, when $q/r$ goes below 1, the filter reduces the gains, placing more
reliance on the accelerometer measurements.
The incorporation of the estimator within the position dynamics can be expressed as
\begin{equation*}
    \begin{aligned}
    &\dot{\hat{y}} = \hat{v}+l_y(y_m-\hat{y}) \\
    &\dot{\hat{v}} = a_m+l_v(y_m-\hat{y})
    \end{aligned}
\end{equation*}
where the Kalman gain vector is $\mathbf{L}=\begin{bmatrix}
l_y & l_v
\end{bmatrix}^T$.
The observations regarding complementary filters done in the context of the rotation
kinematics also apply here.
The accelerometer output successfully tracks rapid fluctuations of the position of the
rocket, but becomes less reliable on the long-term.
On the other hand, the GPS solutions are expected to maintain their mean value, even if
displaying high-frequency disturbances caused by the measurement's inner inaccuracies.
Let us consider $F_y(s)$ and $F_a(s)$ to be the transfer functions from the sensing
results ($y_m$ and $a_m$, respectively) to the estimated position $\hat{y}$;
the closed-loop equations above demonstrate that
\begin{equation*}
    F_{y}(s) = \frac{sl_y + l_v}{s^2+sl_y + l_v},\quad
    F_{a}(s) = \frac{1}{s^2+sl_y + l_v}
\end{equation*}
thus verifying $F_{y}(s) + s^{2}F_{a}(s)=1$.
A trivial analysis of the transfer functions verifies the low- or high-pass
characteristics of the addition terms.

Diving back into the computation of the Kalman gain, the noise sources were considered
to induce covariances equal to $q = 10^{-1}\, [(\text{m/s}^2)^2]$ and $r = 1\,
[\text{m}^2]$.
In such conditions, the Kalman gain is optimized to $\mathbf{L}=\begin{bmatrix}
0.7953 & 0.3162
\end{bmatrix}^T$.
The eigenvalues of the closed loop dynamics are $-0.3976 \pm 0.3976i$, thus verifying
the stability of the derived solutions.

\begin{figure}[ht]
    \centering
    \includegraphics{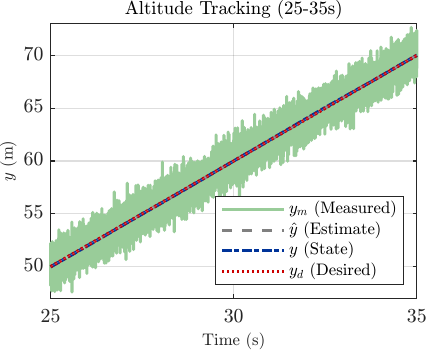}
    \caption{Altitude estimation through the Kalman filter.}
    \label{fig:altitude_estimation.pdf}
\end{figure}

The resulting altitude estimation is presented in Fig.~\ref{fig:altitude_estimation.pdf}, as operated during the 2-D tracking simulations of Section~\ref{sec:simulation}.
The filter is showcased to accurately estimate the true value of the altitude from the highly-oscillatory GPS measurements.
Numerically speaking, the estimation is performed with a standard deviation of $0.0281$ meters from the actual position.

\section{Simulated Experiments}\label{sec:simulation}

\begin{figure}[ht]
    \centering
    \includegraphics[width=1\linewidth]{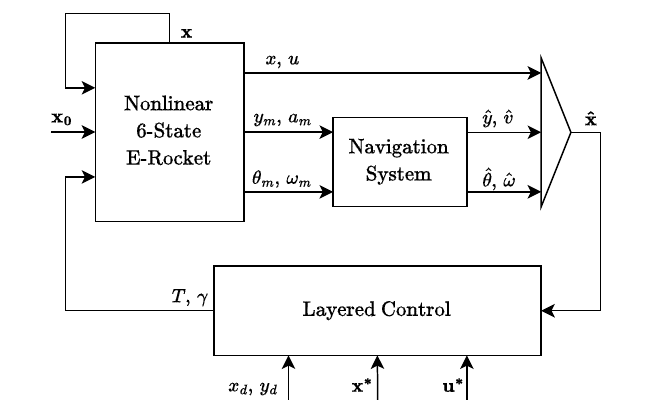}
    \caption{Two-dimensional trajectory tracking setup.}
    \label{fig:2dtt_diagram.pdf}
\end{figure}

The motion within both longitudinal and lateral modes for a modularized and simplified version of the nonlinear model has been tested, showcasing promising results. 
The natural step is to combine both designs such that the e-rocket is able to track two-dimensional steady climbs.
The resulting setup is depicted in Fig.~\ref{fig:2dtt_diagram.pdf}.
The navigation systems are also implemented, allowing for the estimation of the intended states.

\begin{figure*}
    \centering
    \includegraphics[width=.85\linewidth]{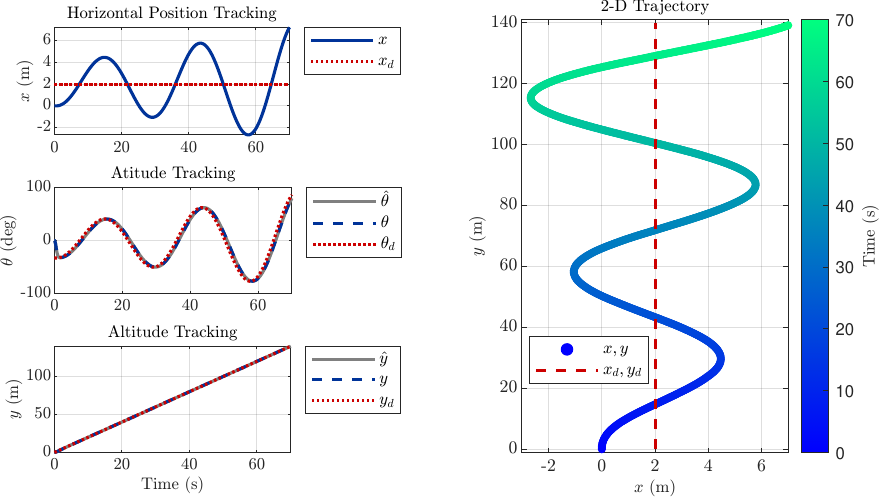}
    \caption{Two-dimensional motion tracking of a vertical 2 m/s ascent at the horizontal position of 2m.}
    \label{fig:2D_response.pdf}
\end{figure*}

The kinematics and dynamics of the vehicle are described by the six-state model of~\eqref{eq:model-6eq-system}, and implemented in Simulink accordingly.
For this new implementation, the sensing conditions of Section~\ref{sec:navigation} are considered, with the noise sources and their covariances set to the values previously defined.
The attitude regulator and altitude controller also retain the previously tuned gains.
The horizontal position controller, on the other hand, has its tuning gain reduced to $k_x=0.01$ to delay the diverging behaviour that we're discussing in a moment.
This means that the system ``rocket + on-board sensors'' has input
$(T,\gamma)$ and output $(x,u,y_m,a_m,\theta_m,\omega_m)$.
The targeted trajectory is maintained as well: the e-rocket starts at rest and at the origin of the inertial frame and targets a steady vertical climb at $x_d=2$m with a constant velocity of $\dot{y}_d=2$ m/s.
The tracking results are presented in Fig.~\ref{fig:2D_response.pdf}.

The presented plots depict successful tracking of the desired altitude, as the e-rocket manages to reach the intended $y$-waypoints at the expected times;
this result aligns with the previous assessments of the altitude controller and estimator.
The pitch angle is also depicted to converge to the desired values, as determined by the outer horizontal position controller.
The latter, however, fails to stabilize the rocket at the desired horizontal position, as the prototype overshoots the reference and oscillates around it until divergence.

\section{Conclusions}\label{sec:conclusions}

The design and implementation of a control and navigation framework for a two-dimensional e-rocket arose valuable insights into the challenges of aerospace engineering.
The architectured solution benefits from its modularity, tackling sub-problems individually and lastly combining them to achieve the mission objectives.
Building the controllers based on Lyapunov theory allowed for global convergence guarantees of the tracking errors.
The state estimators, based on Kalman filtering, protect the implementation from real-world stochasticity.

The conduced simulations revealed strong limitations to the lateral controller, which failed to stabilize the prototype at the desired horizontal position.
These derive from the model idealizations during the design phase, which neglected the presence of drift in the kinematics.
A backstepping solution, as derived for the altitude in Section~\ref{subsec:vertical-position-tracking}, is expected to solve this limitation, as it guarantees the convergence of the tracking error and its derivative to the origin.
These observations highlight the importance of the model's fidelity in the design of control solutions, as commonly-operated simplifications to fully-descriptive models can lead to designing unsuitable controllers.
The implementation of this solution is left for future work.

A nonlinear controllability analysis using tools deriving from Lie algebra can 
provide deeper insights into the system's inherent limitations due to nonlinearities, state coupling, and underactuation constraints.
This suggestion lays the path for future research, where the e-rocket will be modelled in three dimensions and subjected to additional sources of uncertainty, such as aerodynamic effects and mechanical constraints.

\section*{Acknowledgements}

The authors acknowledge the Portuguese Foundation for Science and Technology (FCT) for its financial support via the projects LAETA Base Funding (DOI:10.54499/UIDB/50022/2020). Pedro dos Santos holds a PhD scholarship from FCT (2023.00268.BD).




\begingroup
\footnotesize 
\bibliography{sn-bibliography}
\endgroup


\end{document}